  \documentclass[final,5p,twocolumn]{elsarticle}

  \usepackage{graphicx, color}

\usepackage{amssymb}

\newcommand{\PT}{{$\cal PT$}}

\newcommand{\bg}{{\bf g}}
\newcommand{\br}{{\bf r}}
\newcommand{\ba}{{\bf a}}

\newcommand{\bn}{{\bf n}}

\newcommand{\bm}{{\bf m}}

\def\be{\begin{equation}}
\def\ee{\end{equation}}
\def\bee{\begin{eqnarray}}
\def\ene{\end{eqnarray}}
\def\bes{\begin{subequations}}
\def\ees{\end{subequations}}

\journal{Physics Letters A}

\begin{document}

\begin{frontmatter}



\title{Giant amplification of modes in parity-time symmetric waveguides}


\author{Vladimir V. Konotop$^{1}$, Valery S. Shchesnovich$^{2}$, and Dmitry A. Zezyulin$^{1,*}$}

\address{
 $^1$Centro de F\'{\i}sica Te\'orica e Computacional, and Departamento de F\'{\i}sica, Faculdade de Ci\^encias,
 Faculdade de Ci\^encias,
 Universidade de Lisboa,
Avenida Professor Gama Pinto 2, Lisboa 1649-003, Portugal
\\
$^2$Centro de Ci\^encias Naturais e Humanas, Universidade Federal
do ABC, Santo Andr\'e, S\~ao Paulo 09210-170, Brazil
\\
$^*$Corresponding author: zezyulin@cii.fc.ul.pt}

\begin{abstract}

The combination of the  interference with the amplification of
modes in a waveguide with gain and losses can result in a giant
amplification of the propagating beam, which propagates without
distortion of its average amplitude.  An increase of the gain-loss
gradient by only a  few times results in a magnification  of the
beam by a several orders of magnitude.

\end{abstract}

\begin{keyword}
Optical waveguides \sep Parity-time symmetry \sep Amplification


\end{keyword}

\end{frontmatter}



\section{Introduction}
Amplification of guided optical waves is necessary to compensate for
unavoidable losses along the propagations distance. The amplification is also important in the initiation process for the
pulse generation requiring high intensities with, possibly, very
low intensity inputs. Additionally, the concept of the
gain-guidance, where the optical fibers and planar waveguides with
gain allow for the efficient propagation of low power modes, was
suggested recently~\cite{Siegman} as an alternative to the
conventional index-guiding structures. The phenomenon was observed
in the Nd-doped optical fibers with low refractive index
profiles~\cite{Siegman2} and served as a basic element for the
single-mode optical laser~\cite{Sudesh}. Later on, the idea of
gain-guidance was extended also to the nonlinear
media~\cite{Malomed,ZKK}.

The gain element causes either   the unbounded growth of the
signal intensity or, if nonlinear dissipation becomes dominant,
the convergence to some stationary or pulsating solution (i.e. to
an attractor)~\cite{Akhmediev}. In all   cases the behavior of the
system is fully determined by the values of the system parameters
and does not depend on the characteristics of the input signal
(provided that the input signal belongs to the attractor basin).
Therefore, in the linear systems  one usually considers
distortions of  modes   due to the gain (implemented, say, by
doping), rather than propagation of such modes for  long
distances.

On the other hand,  a large wave intensity in some spatial domains
can be achieved by the constructive interference of modes. In such
amplification scenario, the  increase of the intensity of the
superimposed modes is determined solely by the characteristics
of the  input beam, and as such is limited by  an {\em a priori}
given value.

In this Letter we propose to exploit {\em the
interference  of the gain-guided modes} which, on the one hand, allows for a flexible control of the output intensity by variation
of the input amplitude, and, on the other hand, benefit from the amplifying properties of the medium, thus resulting in a giant
amplification of input beams over relatively  short propagation distances.
Moreover, by a proper design of the dissipation and gain profiles
it is possible to achieve the guidance of modes for an arbitrary
distance, without overcoming {an  {\it a priori} given} intensity
limit, and to realize either single or multichannel guidance
(selected by changing the input beam) inside the same waveguide without changing the waveguide properties.
Below, we focus on the linear case, when the effects of
nonlinearity can be neglected.

Our idea is based  on the use of the special design for  the gain
and loss distribution to assure the parity-time (or \PT) symmetry
of the system. Suggested in the quantum mechanical context~\cite{Bender}, this concept implies spatial (${\cal P}$) and temporal (${\cal T}$) symmetry of the system, which in optics this can be implemented by means of even distribution of the real part of the refractive index of the guiding medium combined with the anti-symmetrically distributed gain and losses~\cite{Muga}. Particular relevance of the \PT-symmetric media for optical applications is justified by the property that  in a definite range of
parameters of such media all linear modes have purely real
propagation constants (which corresponds to the pure real spectrum of the quantum-mechanical Hamiltonian~\cite{Bender}). In other words, this means that {\em all} linear modes propagate without infinite growth or decay, thus allowing for the mode guidance in  media with gain and losses. It is relevant to notice here that  the  ${\cal PT}$ symmetry may be not
sufficient for existence of  purely real spectrum. The
situation when a \PT-symmetric system acquires (for some values of the parameters)   complex eigenvalues (propagation constants)
is usually referred to as spontaneous \PT-symmetry breaking. In this Letter, however, we consider a situation when \PT~symmetry breaking  does not occur.

The setup for the experimental
implementation  of \PT-symmetric  systems with gain and losses using  the wave guiding
structures was proposed in~\cite{Muga}, and later developed in
numerous studies (e.g., for the linear wave propagation
see~\cite{Mustafa,lattice1}). The first experimental studies   of
the optical \PT-symmetric structures were reported
in~\cite{experiemnt1,experiment2}. {In particular,
such phenomena as  spontaneous \PT~symmetry breaking and power
oscillations violating left-right symmetry, have been
experimentally observed.} It was also suggested that appropriately
designed combinations of the gain and losses in a dual core
coupler can be used for the amplification of light
signals~\cite{Malomed1} and for the noise
filtering~\cite{Malomed2}.

\section{Some properties of \PT-symmetric potentials}

We consider  propagation of a paraxial beam governed by the dimensionless field
$q$ obeying the Schr\"odinger equation
 \begin{eqnarray}
\label{CNLS} i {q}_z ={\cal H}q, \qquad {\cal H} =- \nabla^2 +
V(\br) + iW({\bf r}),
\end{eqnarray}
where  $\br=(x,y)$, $\nabla=(\partial_x,\partial_y)$, $r=|{\bf
r}|$, $V(\br)$ describes radially symmetric modulation of the
waveguide refractive index, while $W(\br)$ describes a spatial
distribution of the gain and losses. We focus on the case
\begin{eqnarray}
\label{pt}
V(-\br) = V(\br)\quad\mbox{ and}\quad   W(\br)=-W(-\br),
\end{eqnarray}
 which is a necessary condition
for the operator ${\cal H}$ to be ${\cal PT}$ symmetric. In
particular, we choose~\cite{Bender2}
\begin{equation}
\label{potential} V(r)=r^2, \quad W(\br)=2g_1x + 2 g_2y =
2\,\bg\cdot\br
\end{equation}
where $\bg=(g_1,g_2)= \frac12\nabla W$ characterizes  the gradient
of the gain-loss term having the amplitude $g$: $g^2 = g_1^2 +
g_2^2$. It is relevant to  notice that the beam amplification
reported below, although less pronounced because of the
interference of a  smaller number of modes, can also be observed
in the one-dimensional setting. The two-dimensional setting,
besides giving a considerably stronger amplification,  offers some
additional effects, such as the beam splitting, also considered
below.

For the stationary modes $q(z, \br) = e^{i\beta_\bn
z}\phi_\bn(\br)$, we obtain the eigenvalue problem ${\cal
H}\phi_\bn =-\beta_\bn\phi_\bn$, where $\bn=(n_1,n_2)$ is an
ordered pair of  nonnegative integers identifying the guided
modes. This problem has a pure real spectrum~\cite{Kato}:
$\beta_\bn=-(2n_1+2n_2+2+g^2)$ with the respective eigenfunctions
given by
\begin{eqnarray}
\label{Hermite} \phi_\bn(\br)=
\frac{H_{n_1}(x+ig_1)H_{n_2}(y+ig_2)}{\sqrt{2^{n_1+n_2}n_1!n_2!\pi}}\nonumber\\
\times e^{-\frac{1}{2}[(x+ig_1)^2+(y+ig_2)^2]},
\end{eqnarray}
where $H_n(\xi)$ are the Hermite polynomials. Since all
$\beta_\bn$ are real,  {\em all}  the liner eigenmodes propagate
undistorted.

In addition, $\tilde{\phi}_\bn= \phi_\bn^*$ (the asterisk stands
for the complex conjugation) is an eigenvector of the Hermitian
conjugated operator ${\cal H}^\dag$: ${\cal
H}^\dag\tilde{\phi}_\bn=-\beta_\bn\tilde{\phi}_\bn$. The two sets,
$ \tilde{\phi}_\bn$ and $\phi_\bn$, constitute the left and right
complete bases~\cite{Kato}. The left and right eigenvectors  are
mutually  orthogonal and normalized:
\begin{equation}
(\tilde{\phi}_\bn,\phi_\bm) = \int \phi_\bn(\br)
\phi_\bm(\br)d\br= \delta_{n_1, m_1}\delta_{n_2, m_2},
\end{equation}
where $\delta_{n, m}$ is the Kronecker delta.
 Therefore, for a given input $q_{0}(\br) =
q(0, \br )$, the field evolution can be found  in   the form
\begin{eqnarray}
\label{expansion} q(z,
\br)=\sum\nolimits_{\bn}c_{\bn}e^{i\beta_\bn z}\phi_{\bn}(\br),
\qquad c_{\bn}=\left(\phi_\bn^*,q_0\right).
\end{eqnarray}
However, the right  eigenmodes $\phi_{\bn}(\br)$ [as well as the
left  ones  $\tilde{\phi}_{\bn}(\br)$] are not orthogonal in the
usual sense. Their scalar product reads
\begin{equation}
\hspace{-0.65cm} \left({\phi}_\bn,\phi_\bm\right) =\! \int \!\!
\phi^*_\bn(\br) \phi_\bm(\br)d\br= D_{n_1 m_1}(g_1)D_{n_2
m_2}(g_2),
\end{equation}
 which means
that the optical energy distribution between the modes changes
during the propagation. The energy transfer between the $\bn$-th
and $\bm$-th eigenmodes is described by the functions
\begin{eqnarray}
\hspace{-0.70cm}D_{n m}(g_j) = (\pm
ig_j)^{n+m-2\mu}\,2^{\frac{\kappa}{2}}{e^{ g_j^2}
}\left(\frac{n!}{m!}\right)^{\pm\frac{1}{2}}
L_\mu^{(\kappa)}(-2g_j^2),
\end{eqnarray}
where $L_\mu^{(\kappa)}(\xi)$ are the generalized Laguerre
polynomials,  $L_\mu^{(0)}(\xi)\equiv L_\mu(\xi)$,
 $\kappa = |n-m|$,  $\mu = \min(n,m)$, and the signs ``$+$'' and ``$-$'' stand for the cases $m\geq n$ and $m<n$, respectively.
Since the Laguerre polynomials do not possess  negative
roots~\cite{Szego}, the coefficients $D_{nm}(g_j)\ne 0$, unless
$g_j = 0$.  Since the spectrum $\beta_\bn$ is equidistant, we
conclude from (\ref{expansion}) that the solution $q(z, \br)$ as
well as the energy flow $U(z)=\int |q|^2d\br$ are $\pi$-periodic
with respect to $z$.   We   emphasize that though the equidistant
spectrum simplifies considerably the analysis, it is not necessary
for  a giant amplification induced by the interference of the
gain-guided modes, which can be observed also for other profiles
of the refractive index. {Notice also that energy oscillations are
known to be a  typical  feature of \PT-symmetric systems
\cite{lattice1,experiment2}}.

We intend to  explore the fact that, {the modes which in the
``conservative'' (i.e. index guiding) waveguides have zero
intensity  at the center of the waveguide, for $g^2\neq0$ carry
nonzero power, and moreover,  for a sufficiently large $g^2$ the
respective modes of the \PT-symmetric operator ${\cal H}$ may
acquire  the global  maximum precisely at at $\br=0$.} Moreover,
we  show that in the case of a \PT-symmetric profile with a
sufficiently large $g^2$, the   intensity maximum is shifted to
the mode with  a high quantum number $\bn$, i.e. with a large
value of the propagation constant. This feature  leads to a
dramatically different interference pattern  for such modes,
allowing for a constructive interference of the  maxima of many
different modes precisely at $\br=0$, thus strongly enhancing the
intensity of the beam. At the same time, a destructive
interference of the modes outside the beam center may fully
eliminate all of  the  secondary interference fringes  observable
in the conventional patterns.

\section{Giant amplification}

Let us consider the  input beam of a Gaussian shape,  which is the
lowest eigenmode of a conservative waveguide with the same
refractive index [i.e.  of (\ref{CNLS})--(\ref{potential}) with
$\bg=0$]: $ q_0(r)= \sqrt{U_0/\pi}e^{-r^2/2}, $ where $U_0=
U(0)=\int |q_0|^2d\br$ is the input beam energy flow. A remarkable
fact is that  the respective   solution $q(z, \br)$ can be found
in an explicit form:
\begin{eqnarray}
q = \sqrt{\frac{U_0}{\pi}}\,e^{-i(2+g^2)z-2i\bg\cdot \br\sin^2 z-i
g^2[\sin(4z)/4 - \sin(2z)]}
\nonumber \\
\times e^{2g^2 \sin^2 z }\,e^{-\frac12[(x - g_1\sin(2z))^2+(y -
g_2\sin(2z))^2]}. \label{eq:simple}
\end{eqnarray}
Moreover, formula (\ref{eq:simple}) is a particular case of a more
general family of  explicit solutions to
(\ref{CNLS})--(\ref{potential}) given by \cite{PGV}:
\begin{eqnarray}
q(z, \br) = \sqrt{U_0}\,e^{i\beta_\bn z+i\ba\cdot \br\cos(2z)-i
\ba\cdot\ba\sin(4z)/4}
\nonumber \\
\times e^{2\ba\cdot\bg\sin^2 z-g^2/2
}\,\frac{\phi_\bn\bigl(x-X_1(z), y-X_2(z)\bigr)
}{\sqrt{L_{n_1}(-2g_1^2)L_{n_2}(-2g_2^2)}}, \label{gen_solut}
\end{eqnarray}
where  $\ba=(a_1,a_2)$ is an arbitrary vector, which   {controls
the angle of the incidence of the input beam}, and
$X_{1,2}(z)=a_{1,2}\sin(2z)$ determine the position of the center
of the mode. Indeed, (\ref{eq:simple}) is obtained from
(\ref{gen_solut}) simply by setting $\ba=\bg $ and $\bn=(0, 0)$.
The energy of the beam described by (\ref{gen_solut})  reads $U(z)
= U_0e^{4\ba\cdot\bg\sin^2z}$. In accordance with the  expansion
(\ref{expansion}), $U(z)$ is a $\pi$-periodic function. If the
product $\ba\cdot\bg$ is positive, i.e.  the beam is tilted to
penetrate deeper in the domain with the gain, then $U(z)$
approaches its maxima $U_{max}= U_0e^{4\ba\cdot\bg}$ at
$z_k=\frac{\pi}{2} + k\pi$ and   there is an exponential
amplification of the input amplitude  achieved due to a combined
effect of the gain-loss gradient $\bg$ and a properly chosen
direction of the input beam, i.e. $\ba$. Negative $\ba\cdot\bg$,
on the other hand, results in a periodic attenuation of the beam.

Returning to expression (\ref{eq:simple}) we find that the
amplitude of the solution has the Gaussian shape with the center
oscillating along the  direction indicated by $\bg$. For $0< z
<\pi/2$ the center of the beam is situated in the domain of gain,
and the   energy $U(z)$ grows. The maximal amplification is
achieved  in the point $z=\pi/2$, i.e. when the center of the beam
returns to the origin (where the gain is compensated by losses).
For $\pi/2 < z < \pi$ the beam resides in the lossy domain and
loses the energy. At $z=\pi$ it returns  to the origin and its
energy approaches the lowest value $U_0$. Then the process is
repeated (this  recursive behavior is a consequence of the
\PT~symmetry and resembles     dynamics in a Hamiltonian
system).

\begin{figure}[tb]
\centerline{
       \includegraphics[width=\columnwidth]{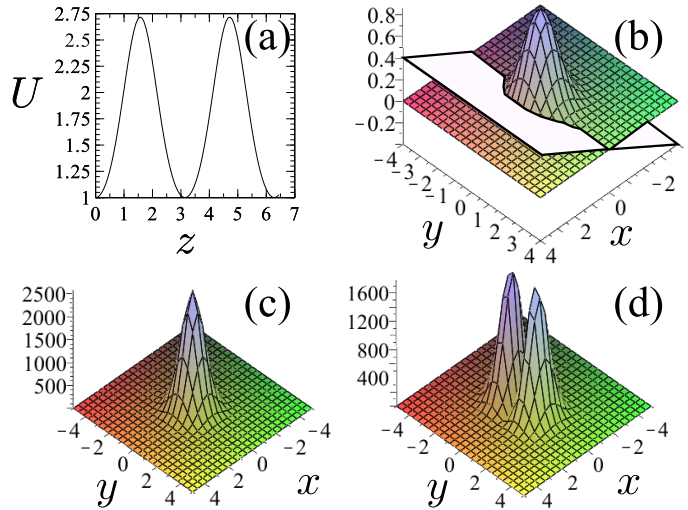}}
\caption{(a) Energy flow {\it vs} propagation distance and (b)
profile $|q(\pi/2, \br)|^2$  for the  solution~(\ref{eq:simple})
with gradient $\bg =(1/2,0)$.  In the panel (b) the gain landscape
$W(\br)$ for $g_1>0$ and $g_2=0$ is schematically shown. (c)
Profile $|q(\pi/2, \br)|^2$ for the solution~(\ref{eq:simple})
with $\bg = (3/2, 0)$. (d) Profile $|q(\pi/2, \br)|^2$ of the
solution (\ref{gen_solut}) with $\bn = (0, 1)$ and gradient $\bg =
\ba = (3/2,0)$. In all panels $U_0=1$.} \label{fig:1}
\end{figure}

To illustrate the above findings we focus, on the case of
$\bg=(g_1,0)$. At first, we consider  the particular case given by
(\ref{eq:simple}). The beam energy  flow $U(z)$ and the beam
profiles at the maximal amplification point are shown in
Fig.~\ref{fig:1}~(a)-(c). Figure~\ref{fig:1} reveals several
important features. First, even a weak increase of the gradient
results in a giant amplification of the beam intensity as it is
seen from   panels (b) and (c) of Fig.~\ref{fig:1}. Whereas these
two panels  are obtained for the same Gaussian input beam, an
increase of the gradient amplitude $g$  by three times [from
$g={1}/{2}$ in panel (b) to $g={3}/{2}$ in panel (c)] results in
magnification of the peak intensity by about  $3\times 10^3$ times
[the peak intensity is about 0.8 in panel (b) against
approximately $2.5\times 10^3$ in panel (c)]. Second,   both in
panels (b) and (c) the maximal beam intensity is achieved right at
the center of the waveguide, although many eigenmodes are excited
and, moreover, the maximal portion of the energy is concentrated
in a higher mode. To illustrate the last statement, in
Table~\ref{tbl:1} we display the coefficients of the expansion
(\ref{expansion}) of the input Gaussian beam $q_0(r)=
\sqrt{U_0/\pi}e^{-r^2/2}$ over the eigenmodes $\phi_\bn$:
\begin{equation}
c_\bn = {\sqrt{U_0 / (2^{n_1+n_2}n_1!n_2!)
}\,\,(ig_1)^{n_1}(ig_2)^{n_2}e^{g^2/4}}.
\end{equation}
 \begin{table}[tb]
  \centering
  \caption{Absolute values of the coefficients  $c_{(n_1, 0)}$   for $U_0=1$, $g_2=0$ and different $g_1$ [for $n_2\neq0$ all  $c_{(n_1, n_2)}=0$].}
  \begin{tabular}{ccccccccccccc} \\ \hline
    $n_1$ & 0 & 1 & 2 &3 &4\\\hline
    $g_1=1/2$    & 1.06  & 0.37 & 0.09 & 0.02 & 0.003\\
    $g_1=3/2$    & 1.76  & 1.86 & 1.40 & 0.85 & 0.45 \\\hline
  \end{tabular}
  \label{tbl:1}
\end{table}
From Table~\ref{tbl:1} one can see,  that for $\bg =(3/2,0)$ the
most  excited mode, i.e.  the one   having the largest coefficient
$c_\bn$,  is the mode corresponding to $n_1=1$ (rather than  the
mode with  $n_1=0$). With further increase of the gradient $g_1$
the maximal amplitude is shifted towards     higher modes.

The obtained results do not exhaust all possibilities of our
setup. In particular,  for  the input beam with the shape of the
first Gauss-Hermite function,   the solution (\ref{gen_solut})
yields giant amplification of the split beam, as shown in
Fig.~\ref{fig:1}~(d). Even more sophisticated profiles of the
amplified beam can be obtained by variation of the input beam,
i.e. by changing $\bn$, or by using the incidence angle ``rotated"
with respect to the gain-loss gradient (i.e. using a different
relation between $\bg$ and $\ba$).

\section{Conclusion}

To conclude, we have proposed  a physical setup where a giant but
controllable, i.e. limited by the amplitude,  amplification of a guided beam can be achieved
during the propagation. Such amplification occurs due to the
interference of the multiple gain guided modes. The latter is possible due to the reality of the propagation constants of all modes, on the one hand, and on the other hand due to the energy transfer among the modes which results in exciting many modes, practically by arbitrary input beam. The phenomenon is determined by the
parameters of both the system and the input beam, thus allowing
for an efficient managing over the output beam amplitude and, to
some degree, over the output beam shape. This last property makes the configurations particularly relevant for the practical purposes whenever it is necessary not only to enhance of an input beam until some given amplitude but also change this limit in the course of the experiment by managing only the input field. We also notice that further studies, including different \PT-symmetric profiles  allowing for the spontaneous symmetry breaking or  nonlinearity, which may become relevant at high intensities of the phenomenon are of interest.

\section*{Acknowledgements}
The authors acknowledge support of the FCT grants
PTDC/FIS/112624/2009, PEst-OE/FIS/UI0618/2011, and
SFRH/BPD/64835/2009, as well as support of FAPESP and CNPq of
Brasil.


\begin{thebibliography}{99}

\bibitem{Siegman}
{ A. E. Siegman, J. Opt. Soc. Am. A {\bf 20}, 1617 (2003).}

\bibitem{Siegman2} A. E. Siegman, Y. Chen, V. Sudesh, M. C. Richardson, M. Bass, P. Foy, W. Hawkins, and J. Ballato,
Appl. Phys. Lett. {\bf 89}, 251101 (2006).

\bibitem{Sudesh} V. Sudesh, T. McComb, Y. Chen, M. Bass, M. Richardson, J. Ballato, A.E. Siegman,
 Appl. Phys. B {\bf 90}, 369 (2008).

\bibitem{Malomed} C.-K. Lam, B. A. Malomed, K. W. Chow, and P. K. A. Wai,
Eur. Phys. J. -- Special Topics {\bf 173}, 233  (2009).

\bibitem{ZKK}  D. A. Zezyulin, Y. V. Kartashov, and V. V. Konotop, Opt. Lett. {\bf 36}, 1200 (2011).

\bibitem{Akhmediev} N. Akhmediev and A. Ankiewicz, eds., \textit{Dissipative Solitons: From Optics to Biology and Medicine} (Springer, Berlin, 2008).

\bibitem{Bender}
C. M. Bender and S. Boettcher, Phys. Rev. Lett. {\bf 80}, 5243
(1998).

\bibitem{Muga} A. Ruschhaupt, F. Delgado, and J. G. Muga, J. Phys. A: Math. Gen. {\bf 38}, L171  (2005).


\bibitem{Mustafa} A. Mostafazadeh and F. Loran, Europhys. Lett. {\bf 81},   10007 (2008).

\bibitem{lattice1}
K. G. Makris, R. El-Ganainy, D. N.  Christodoulides, and Z. H.
Musslimani, Phys. Rev. Lett. {\bf 100}, 103904 (2008).

\bibitem{experiemnt1} A. Guo, G. J. Salamo, D. Duchesne, R. Morandotti, M. Volatier-Ravat,  V. Aimez, G. A. Siviloglou and D. N. Christodoulides, Phys. Rev. Lett. {\bf 103}, 093902 (2009).


\bibitem{experiment2}
C. E. R\"uter, K. G.  Makris, R. El-Ganainy, D. N.
Christodoulides,  M. Segev, and   D. Kip,  Nature Phys. {\bf 6},
{192} {(2010)}.

\bibitem{Malomed1} B. A. Malomed, G. D. Peng, and P. L. Chu,  Opt. Lett. {\bf 21},
 330
 (1996).

\bibitem{Malomed2} P. L. Chu, G. D. Peng, B. A. Malomed, H. Hatami-Hanza, and I. M. Skinner
Opt. Lett. {\bf 20}, 1092
(1995).



\bibitem{Bender2} C. M. Bender  and H. F Jones,
J. Phys. A: Math. Theor. {\bf 41}, 244006  (2008).


\bibitem{Kato} T. Kato, \textit{Perturbation theory for linear operators} (Springer-Verlag, 1980).

\bibitem{Szego}  G. Szeg\"o, \textit{Orthogonal Polynomials}  (American Mathematical Society, Providence, RI, 1939).

\bibitem{PGV} J. J. Garc\'{\i}a-Ripoll, V. M. P\'erez-Garc\'{\i}a, and V. Vekslerchik,
Phys. Rev. E {\bf 64}, 056602 (2001).

\end{thebibliography}
\end{document}